\documentstyle [12pt]{article}
\title{Natural Mass Matrices}
\author{R. D. Peccei and K. Wang
\\ Dept. of Physics, University of California, Los Angeles
\\ Los Angeles, California 90024}
\date{}
\def\dfrac{\displaystyle\frac}

\def\gsim{\mathrel{\raise.3ex\hbox{$>$\kern-.75em\lower1ex\hbox{$\sim$}}}}
\def\lsim{\mathrel{\raise.3ex\hbox{$<$\kern-.75em\lower1ex\hbox{$\sim$}}}}

\begin{document}

\maketitle

\begin{abstract}
We introduce the idea of natural mass matrices, an organizing principle useful
 in the search for  GUT scale quark mass matrix patterns that are consistent
with known CKM constraints and quark mass eigenvalues. An application of this
idea is made in the context of SUSY GUTs and some potentially ``successful''
GUT scale mass patterns are found. The CKM predictions of these patterns are
presented and some relevant strong CP issues are discussed.
\end{abstract}

\vspace{4.5cm}
\begin{flushleft}
$\quad$UCLA/95/TEP/29

$\quad$August 1995
\end{flushleft}
\pagebreak

\section{\normalsize \bf Introduction}
\setcounter{equation}{0}

It has been a theoretical quest for nearly 20 years to devise interesting mass
matrix patterns which could provide sound predictions for the CKM matrix and
quark mass eigenvalues~\cite{FRITZ}. Recently, these efforts have centered on
constructing patterns at the GUT scale in SUSY
theories~\cite{DHR}~\cite{ARDHS}. Most of these attempts, although quite
successful, have failed to produce
results in complete agreement with precise low energy
data(LED)\cite{PECCEI_talk}. Part of the difficulty lies in the fact that one
has to rely upon ``guesses'' at the GUT scale which are then extrapolated down
to the weak scale, with the hope that the mass matrices so obtained will give
rise to acceptable fits to the LED. A somewhat more promising approach perhaps,
would be to reverse the process: constructing LED-consistent mass matrices at
some low energy scale and then evolve them upwards to see if interesting GUT
patterns would emerge. However, since there exists considerable arbitrariness
in this construction, one typically has to be content with studing only certain
special cases~\cite{AN}\cite{RRR} often guided merely by simplicity. Here, we
would like to suggest an organizing principle which may be helpful. This is the
idea of natural mass matrices, which severely restricts the aforementioned
arbitrariness in the mass pattern construction, thereby allowing a search for
viable GUT patterns more systematically and efficiently.

\medskip

This paper is organized as follows: in Sec.~2,  we summarize our present
knowledge of the CKM parameters and quark masses; in Sec.~3, we introduce the
idea of natural mass matrices along with our method of
mass-matrix-parametrization that facilitates its implementation; in Sec.~4, we
present some Hermitian GUT scale mass patterns that emerge from this approach
and their CKM predictions; in Sec.~5, we discuss issues connected with
Hermiticity breakdown and associated problems with strong CP violation; finally
in Sec.~6, we present our conclusions.

\bigskip

\section{\normalsize \bf CKM and Quark Masses: a Summary of Present Status}
\setcounter{equation}{0}

\begin{flushleft}
\it{2.1.   CKM Matrix}
\end{flushleft}
\rm

In its standard form~\cite{PDG}, the Cabibbo-Kakayashi-Maskawa (CKM) matrix is
\small
\begin{equation}
[CKM]_{s} = \left( \begin{array}{ccc}
 c_{1}c_{3}   &  s_{1}c_{3}     &  s_{3}e^{-i\delta}  \\
 -s_{1}c_{2}-c_{1}s_{2}s_{3}e^{i\delta} &
c_{1}c_{2}-s_{1}s_{2}s_{3}e^{i\delta}     &  s_{2}c_{3}  \\
 s_{1}s_{2}-c_{1}c_{2}s_{3}e^{i\delta} &
-c_{1}s_{2}-s_{1}c_{2}s_{3}e^{i\delta}     &  c_{2}c_{3}
 \end{array} \right) \;\; . \label{CKM}
\end{equation} \normalsize Taking into account of the experimental hierarchy in
the mixing angles, one can write \begin{eqnarray}
s_{1}  & \equiv & \sin \theta_{1} \equiv \lambda \simeq 0.22 \;\; ,  \nonumber
\\
s_{2}  & \equiv & \sin \theta_{2} \equiv A \lambda^{2} \;\; ,  \nonumber \\
s_{3}  & \equiv & \sin \theta_{3} \equiv A \sigma \lambda^{3} \;\; ,
\label{CKM_No}
\end{eqnarray} with $A, \; \sigma$ being parameters roughly of $O(1)$. The CKM
matrix then takes the Wolfenstein~\cite{WOLF} form~\footnote{Wolfenstein uses
the parameters $\rho, \eta$ instead of $\sigma, \delta$. They are related by
$\sigma e^{-i\delta} \equiv \rho - i\eta$.} \small \begin{eqnarray}
 [CKM]_{s} & = & \left( \begin{array}{ccc}
 1 - \frac{1}{2}\lambda^{2} - \frac{1}{8}\lambda^{4} &
 \lambda     &
 A \sigma \lambda^{3} e^{-i\delta}  \\
 - \lambda &
 1 - \frac{1}{2} \lambda^{2} - (\frac{1}{2}A^{2} + \frac{1}{8})\lambda^{4} &
 A \lambda^{2} \\
 A \lambda^{3}(1- \sigma e^{i\delta}) &
 -A\lambda^{2} + \frac{1}{2}A\lambda^{4} &
 1-\frac{1}{2}A^{2}\lambda^{4}
 \end{array} \right)  \nonumber  \\ \nonumber \\ \nonumber \\
& &  + \, O(\lambda^{5})  \, . \label{CKM_W}
\end{eqnarray} \normalsize

\smallskip

The parameter $A$ is fixed by $V_{cb}$ which, in turn, is determined from
semileptonic B decays. The most recent analysis from CLEO data~\cite{Stone}
gives,
\begin{equation}
|V_{cb}|  =  0.0378 \pm 0.0026 \; \leftrightarrow \; A = 0.78 \pm 0.05 \;\; .
\label{A}
\end{equation}

The parameter $\sigma$ is fixed by the ratio $|V_{ub}|/|V_{cb}|$. This in turn
can be extracted from a study of semileptonic B decays near the end point
region of the lepton spectrum, where $b \rightarrow c$ quark transitions are
forbidden. The most recent analysis~\cite{Stone} gives,
\begin{equation}
|V_{ub}|/|V_{cb}|  =  0.08 \pm 0.02 \; \leftrightarrow \; \sigma = 0.36 \pm
0.09 \;\; . \label{sigma}
\end{equation}

The phase $\delta$ (or the parameter $\eta$) can be gotten by combining the
measurements of the $\epsilon$  parameter in $K - \overline{K}$ mixing and
those of $\Delta m_{B_{d}}$ in $B_{d} - \overline{B}_{d}$ mixing with the value
of
$|V_{ub}|/|V_{cb}|$.  A recent analysis~\cite{PW} gives an allowed region in
the $\rho - \eta$ plane roughly specified by the ranges \begin{equation}
\eta  \simeq  [0.2, \; 0.5]  \;\;\; , \;\;\; \rho  \simeq  [-0.4, \; 0.3] \;\;
\label{rho_eta}
\end{equation} with a corresponding CKM phase \begin{equation}
\delta \simeq  [45^{0}, \; 158^{0}] \;\; .
\end{equation}

\begin{flushleft}
\it{2.2.   Quark Masses}
\end{flushleft}
\rm

For the purpose of calculating quark mass ratios and their RG scaling, it is
convenient to express all quark masses as running masses at some common energy
scale. We shall choose this scale here to be the mass of the top quark $m_{t}$.
We summarize below what is known about the quark masses and then extrapolate
all the results to the scale $m_{t}$. From the recent ``discovery'' papers on
the
top quark~\cite{CDF}, one infers a value for the physical mass $m^{Phys}_{t}$,
which is related to the running mass by \[ m_{t}(m_{t}) =
\frac{m^{Phys}_{t}}{1+\frac{4}{3\pi}\alpha_{s}(m_{t})} \;\; . \] These
results~\cite{CDF} suggest that \[ m_{t}(m_{t})  \simeq  (175 \pm 15) \, GeV .
\] For medium heavy quarks, the analyses of charmonium and bottomonium
spectra~\cite{GL} give \begin{eqnarray*}
m_{c}(m_{c}) & = & (1.27 \pm 0.05) \, GeV \;\; , \\
m_{b}(m_{b}) & = & (4.25 \pm 0.10) \, GeV \;\; .
\end{eqnarray*} Finally, for light quarks, current algebra analyses~\cite{GL}
give the following values for the masses at a scale of $1 GeV$,
\begin{eqnarray*}
m_{u}(1 GeV) & = & (5.1 \pm 1.5) \, MeV \;\; , \\
m_{d}(1 GeV) & = & (8.9 \pm 2.6) \, MeV \;\; , \\
m_{s}(1 GeV) & = & (175 \pm 55) \, MeV \;\; .
\end{eqnarray*} These mass values are individually uncertain to $O(30\%)$ but
are rather better constrained~\cite{KM}~\cite{Leut} by the current algebra
relation \begin{equation}
(\frac{m_{u}}{m_{d}})^{2}+\frac{1}{Q^{2}}(\frac{m_{s}}{m_{d}})^{2} = 1 \;,
\;\;\; \mbox{with} \;\; Q = 24 \pm 1.6 \; . \label{light_q}
\end{equation}

\smallskip

The RG scaling of the medium heavy and light quark masses to $m_{t}$ are
calculated to 3-loops in Ref.~\cite{ARDHS}, with the result sensitive to the
precise value of the strong coupling constant $\alpha_{s}(M_{Z})$. Using
$\alpha_{s}(M_{Z}) = 0.117 \pm 0.05$ and expressing all quark mass ratios in
terms of the small parameter $\lambda \simeq 0.22$, the above results allow one
to write the diagonal quark mass matrices as

\begin{eqnarray}
M_{u}^{diag}(m_{t}) & = & 175 \, GeV \left( \begin{array}{ccc}
 \xi_{ut}\lambda^{7} &  0      & 0 \\
 0      &  \xi_{ct}\lambda^{4} & 0 \\
 0      &  0      &  \xi_{tt}
\end{array} \right) \nonumber \\ \nonumber \\
& \equiv & m_{t}(m_{t}) \, \tilde{M}_{u}^{diag} \; , \label{Mudiag}
\end{eqnarray} where $\; \xi_{ut} = 0.49 \pm 0.15$,  $\; \xi_{ct} = 1.46 \pm
0.13$,  $\; \xi_{tt} = 1 \pm 0.09$; and
\begin{eqnarray}
M_{d}^{diag}(m_{t}) & = & 2.78 \, GeV \left( \begin{array}{ccc}
 \xi_{db}\lambda^{4} &  0      & 0 \\
 0      &  \xi_{sb}\lambda^{2} & 0 \\
 0      &  0      & \xi_{bb}
\end{array} \right) \nonumber \\ \nonumber \\
& \equiv & m_{b}(m_{t}) \, \tilde{M}_{d}^{diag} \; , \label{Mddiag}
\end{eqnarray} where $\; \xi_{db} = 0.58 \pm 0.18$,  $\; \xi_{sb} = 0.55 \pm
0.18$, $\; \xi_{bb} = 1 \pm 0.05$.

\bigskip

\section{\normalsize \bf Natural Mass Matrices}
\setcounter{equation}{0}

\begin{flushleft}
\it{3.1.   A Heuristic Two-Generation Example -- the Notion of Naturalness}
\end{flushleft}
\rm

To introduce the idea of natural mass matrices, we consider first the simple
two-quark-generation case, assuming that these mass matrices are Hermitian. A
general 2$\times$2 Hermitian mass matrix for the first two quark families can
always be rewritten, after some trivial phase redefinitions of the quark
fields, as some real symmetric matrix. Two such matrices for the u-quarks and
d-quarks:
$M_{u} \equiv m_{c}\, \tilde{M}_{u}$, $M_{d} \equiv m_{s}\,\tilde{M}_{d}$, in
turn can be diagonalized by some orthogonal matrices $O_{u}$ and $O_{d}$,
resulting in a Cabbibo-quark-mixing matrix $C$. One has\footnote{Note that
$\xi_{uc} = \xi_{ut}/(\xi_{ct}\lambda) = (1.53 \pm 0.49)$ while $\xi_{ds} =
\xi_{db}/\xi_{sb} = (1.05 \pm 0.47)$.}  \begin{eqnarray}
O^{T}_{u} \tilde{M}_{u} O_{u} & = & \tilde{M}_{u}^{diag} \equiv \left(
\begin{array}{cc}
\xi_{uc}\lambda^{4} &  0   \\
 0   & 1
\end{array} \right) \; , \nonumber \\
O^{T}_{d} \tilde{M}_{d} O_{d} & = & \tilde{M}_{d}^{diag} \equiv \left(
\begin{array}{cc}
 \xi_{ds}\lambda^{2} &  0   \\
 0   & 1
\end{array} \right)  \label{MM2}
\end{eqnarray} and \begin{equation}
C = O^{T}_{u} O_{d} = \left( \begin{array}{cc}
 \cos \theta_{C} &  \sin \theta_{C}   \\
- \sin \theta_{C}   &   \cos \theta_{C}
\end{array} \right) \;\; . \label{C2}
\end{equation}

In the above, $\lambda \equiv \sin \theta_{C}$ as before and the matrices
$O_{u}$,  $O_{d}$ have the same form as the matrix $C$ with angles
$\theta_{u}$, $\theta_{d}$ instead of $\theta_{C}$. We notice that it follows
from
Eq.(\ref{C2}) that $\theta_{C}=\theta_{d}-\theta_{u}$. Moreover, since the
matrix $C$ is invariant under the changes \[ O_{d} \rightarrow O \, O_{d} \; ;
\;\;\; O_{u} \rightarrow O \, O_{u} \; , \] where $O$ is some arbitrary
orthogonal matrix, we see that $ \tilde{M}_{u}$ and $ \tilde{M}_{d}$ are fixed
only up to a common
similarity transformation \[  \tilde{M}_{u} \leftrightarrow O^{T}
\tilde{M}_{u} \, O \; ; \;\;\;  \tilde{M}_{d} \leftrightarrow O^{T}
\tilde{M}_{d} \, O \; . \]

Because of this freedom and since $\theta_{C} \sim \lambda << 1$, we can always
arrange to have both $\theta_{d} << 1$ and $\theta_{u} << 1.$~\footnote{As will
become apparent below, this restriction need not to be imposed separately for
it is in conformity with our naturalness requirement. It is done here for ease
in
the discussion that follows.} We can now contemplate three different options
for the angles $\theta_{u}$ and $\theta_{d}$: \begin{equation}
\begin{array}{cll}
(i) & \sin \theta_{d} \sim \lambda \; , &  \sin \theta_{u} \sim \lambda \; ; \\
(ii) & \sin \theta_{d}  \sim  \lambda \; , &  \sin \theta_{u} \lsim \lambda^{2}
\; ; \\
(iii) & \sin \theta_{d}  \lsim  \lambda^{2} \; , &  \sin \theta_{u} \sim
\lambda \; . \label{twogen}
\end{array}
\end{equation} Upon examining the expressions for the mass matrices
$\tilde{M}_{u}$ and $\tilde{M}_{d}$:  \begin{eqnarray}
\tilde{M}_{u} & = &  O_{u} \tilde{M}_{u}^{diag} O^{T}_{u} \simeq \left(
\begin{array}{cc}
 \xi_{uc}\lambda^{4}+ \sin^{2} \theta_{u} &  \sin \theta_{u}   \\
 \sin \theta_{u}    & 1
\end{array} \right) \; , \nonumber \\
\tilde{M}_{d} & = &  O_{d} \tilde{M}_{d}^{diag} O^{T}_{d} \simeq \left(
\begin{array}{cc}
 \xi_{ds}\lambda^{2}+ \sin^{2} \theta_{d} &  \sin \theta_{d}   \\
 \sin \theta_{d}    & 1
\end{array} \right) \; , \label{twogenexp}
\end{eqnarray}  one sees that if options $(i)$ and $(iii)$ were to hold, one
would require a severe fine-tuning of the matrix element
$[\tilde{M}_{u}]_{11}$, forcing $[\tilde{M}_{u}]_{11} \simeq
([\tilde{M}_{u}]_{12})^{2}$ to arrive at
the large $m_{u}/m_{c} \sim \lambda^{4}$ experimental hierarchy. Such a
fine-tuning appears to be unnatural. For the two-generation case, the natural
mass matrices have the form \begin{equation}
\tilde{M}_{u} \simeq \left( \begin{array}{cc}
\alpha'_{u}\lambda^{4} &  \alpha_{u}\lambda^{2}   \\
\alpha_{u}\lambda^{2}     & 1
\end{array} \right) \;\; ; \;\; \tilde{M}_{d}  \simeq  \left( \begin{array}{cc}
\alpha'_{d}\lambda^{2} &  \alpha_{d}\lambda   \\
\alpha_{d}\lambda      & 1
\end{array} \right) \; ,
\end{equation} corresponding to option $(ii)$ in which \[ \begin{array}{ll}
\sin \theta_{u} =  \alpha_{u} \lambda^{2} \;, & \alpha'_{u}- \alpha^{2}_{u} =
\xi_{uc} \; ; \\
\sin \theta_{d} = \alpha_{d} \lambda \; , & \alpha'_{d}- \alpha^{2}_{d} =
\xi_{ds} \;
\end{array} \] and, \[ \alpha_{d} - \alpha_{u}\lambda \simeq 1 \; . \] With the
parameters $\alpha$ and $\alpha'$ of $O(1)$, one gets the observed hierarchy
(i.e. the $\xi$'s being of $O(1)$) without any need for fine-tunings.

\medskip

\begin{flushleft}
\it{3.2.   Three-Generation Extension}
\end{flushleft}
\rm

It is straightforward, though somewhat tedious, to extend the idea of natural
mass matrices to the three-generation case. To facilitate its implementation
however, we need to introduce a convenient parametrization of the mass matrices
based on a perturbative expansion in ``$\lambda$'' . The procedure we shall
adopt is analogous to, but a slight generalization of, a method employed by
Ramond, Roberts and Ross~\cite{RRR}. The benefits of our generalization, aside
from allowing a search for natural mass patterns, also include the flexibility
of simultaneous adjustments of matrix elements in $M_{u}$ and $M_{d}$ (useful
in imposing ``0''s or arranging for equalities among them).

Consider some general 3$\times$3 Hermitian mass matrices $M_{u} \equiv
m_{t}(m_{t}) \tilde{M}_{u}$ and $M_{d} \equiv m_{b}(m_{t}) \tilde{M}_{d}$.
These matrices can be diagonalized by some unitary matrices $U$ and $D$,
resulting in
a CKM-quark-mixing matrix $[CKM]$: \begin{eqnarray}
\tilde{M}_{u} & = & U \, \tilde{M}_{u}^{diag} \, U^{\dagger} \; , \label{mu3}
\\
\tilde{M}_{d} & = & D \, \tilde{M}_{d}^{diag} \, D^{\dagger} \; , \label{md3}
\\
{[CKM]} & = & U^{\dagger}D \; . \label{ckm3}
\end{eqnarray} To proceed, it is useful to make two observations:

(1) As in the two-generation case, if we change $U \rightarrow NU$ and $D
\rightarrow ND$ (where $N$ is some arbitrary unitary matrix), the matrix
$[CKM]$ remains unchanged. The matrices $\tilde{M}_{u}, \, \tilde{M}_{d}$ are
thus
unique up to an arbitrary(but common) unitary transformation: $\tilde{M}_{u}
\leftrightarrow N^{\dagger}(\tilde{M}_{u})N, \, \tilde{M}_{d} \leftrightarrow
N^{\dagger}(\tilde{M}_{d})N$. As a result of $[CKM] \simeq 1$, this
arbitrariness allows us to focus only on ``small'' transformations, i.e. $U
\simeq 1, \; D \simeq 1$. Furthermore, in the particular case where
$N=\phi_{L}$, $\phi_{L}$ being  some phase matrix, the induced changes in
$\tilde{M}_{u}$ and $\tilde{M}_{d} \; (\tilde{M}_{u} \rightarrow \phi_{L}
\tilde{M}_{u} \phi^{\dagger}_{L} \; ; \tilde{M}_{d} \rightarrow \phi_{L}
\tilde{M}_{d} \phi^{\dagger}_{L})$ can be absorbed by a trivial quark field
phase redefinition.

(2) So far as constructions of the mass matrices $\tilde{M}_{u}$ and
$\tilde{M}_{d}$ out of the matrices $\tilde{M}_{u}^{diag}, \;
\tilde{M}_{d}^{diag}$ and $[CKM]$ are concerned, a change of $[CKM] \rightarrow
\phi^{\dagger}_{u} \, [CKM]  \, \phi_{d} \Longleftrightarrow D \rightarrow D
\phi_{d} \, ;  U \rightarrow U \phi_{u}$, with the $\phi$'s some arbitrary
phase matrices, is inconsequential due to the fact that $\phi_{u}
\tilde{M}_{u}^{diag}
 \phi^{\dagger}_{u} = \tilde{M}_{u}^{diag}$ and $ \phi_{d} \tilde{M}_{d}^{diag}
\phi^{\dagger}_{d} = \tilde{M}_{d}^{diag}$.

\smallskip

Based on these observations, one is always justified working with a specific
form of the CKM matrix. Thus one can write \[ U^{\dagger}D = [CKM] \equiv
\phi^{\dagger}_{u} \, [CKM]_{s}  \, \phi_{d} \] with \[ D \equiv
\phi^{\dagger}_{L} D_{s} \phi_{d} \; ; \; U \equiv \phi^{\dagger}_{L} U_{s}
\phi_{u} \;. \] The phase matrices $\phi_{L}, \; \phi_{d}$ can be further
chosen so as to render $D_{s}$ to have the same form as the matrix $[CKM]_{s}$
of
Eq.(\ref{CKM}), and it follows that $U^{\dagger}_{s} D_{s} =
[CKM]_{s}$. (Notice that having chosen $D_{s}$ to be of the same form as the
matrix $[CKM]_{s}$, there is now no more freedom to redefine $U_{s}$ and
$U_{s}$, in fact, will not have quite the standard CKM form.) The
mass matrices $\tilde{M}_{u}$,  $\tilde{M}_{d}$ constructed as \begin{eqnarray}
\tilde{M}_{u} & = & U_{s} \, \tilde{M}_{u}^{diag} \, U^{\dagger}_{s} \; ;
\nonumber \\
\tilde{M}_{d} & = & D_{s} \, \tilde{M}_{d}^{diag} \, D^{\dagger}_{s} \label{MM}
\end{eqnarray} are still perfectly general for our considerations.\footnote{The
matrices $\tilde{M}_{u}$ and  $\tilde{M}_{d}$ in Eq.(\ref{MM}) can still be
changed by an overall common unitary transformation. It is possible to remove
this remaining freedom by adopting a (slightly) unconventional parametrization
of the CKM matrix. Defining $\tilde{M}_{u}$ and  $\tilde{M}_{d}$ as in
Eqs.(\ref{mu3}) and (\ref{md3}), one has always the freedom to choose $D$ to be
of the standard CKM form $D=D_{s}$, so that the matrix $\tilde{M}_{d}$ is
described by 6 real variables and 1 phase. However, then there is no further
freedom of redefinition, hence $U$ being a general 3$\times$3 Hermitian matrix
must involve 6 real variables and 3 phases. It is easy to show that this more
general parametrization corresponds to a definition of the CKM matrix (in the
standard form) $[CKM]_{s}= U^{\dagger}D_{s}\Phi$, with $\Phi$ being a diagonal
phase matrix containing 2 arbitrary phases. We find it more convenient to take
$\Phi=1$ and have the matrices $\tilde{M}_{u}$ and  $\tilde{M}_{d}$
undetermined by an overall common unitary transformation.} Having
established their generality, we shall drop the subscript ``s'' in the matrices
$[CKM]_{s}$, $D_{s}$ and $U_{s}$ hereafter for notational brevity.

\smallskip

With these preliminaries out of the way, it is straightforward to give explicit
parametrizations for the matrices $U$ and $D$. For these purposes, it is useful
to note that, in its standard form, the matrix $[CKM]$ can be expressed as
\small  \begin{eqnarray*}
[CKM] & = & C_{2} \, \Delta \, C_{3} \, \Delta^{\dagger} C_{1} \nonumber \\
\nonumber \\
& = &  \left( \begin{array}{ccc}
1  & 0 & 0 \\
0 & c_{2} & s_{2} \\
0 &  -s_{2} & c_{2}
\end{array} \right) \Delta \left( \begin{array}{ccc}
c_{3}  & 0 & s_{3} \\
0 & 1 & 0 \\
-s_{3} & 0  & c_{3}
\end{array} \right) \Delta^{\dagger} \left( \begin{array}{ccc}
c_{1}  & s_{1} & 0 \\
-s_{1} & c_{1} & 0 \\
0      &    0  & 1
\end{array} \right)
\end{eqnarray*}  \normalsize with \[ \Delta \equiv  \left( \begin{array}{ccc}
1  & 0 & 0 \\
0  & 1 & 0 \\
0  & 0 & e^{i\delta}
\end{array} \right) \;\; . \] Because $D$, by assumption, also takes the CKM
form, we can likewise write  \begin{equation}
D =  C_{2d} \, \Delta_{d} \, C_{3d} \, \Delta^{\dagger}_{d} C_{1d} \; .
\label{D}
\end{equation} Here the matrices $C_{id}$ and $\Delta_{d}$ are defined
analogously to $C$ and $\Delta$, except they involve some new angles
$\theta_{id}  \; (i=1,2,3)$ and $\delta_{d}$ . It is convenient, in addition,
to define three more orthogonal matrices $C_{iu}$ (of the same form as the $
C_{i}$'s) involving angles  $\theta_{iu} \,  (i=1,2,3)$ and satisfying the
relations \[ C^{T}_{iu} C_{id} = C_{i} \; \mbox{(no sum over the index ``i'')}
\; . \] In analogy to the two-generation case, the angles $\theta_{iu}, \,
\theta_{id}$ and $\theta_{i}$ obey \begin{equation}
\theta_{i} =  \theta_{id} - \theta_{iu} \;\; (i=1,2,3) \;\; . \label{theta_i}
\end{equation} Using these definitions, the matrix $U$ is easily seen to be
\begin{eqnarray}
U & = & D \,{[CKM]}^{\dagger} \nonumber \\
& = & \{C_{2u}\} \, \{ C_{2} \, (\Delta_{d} \, C_{3u} \, \Delta^{\dagger}_{d})
\, C^{\dagger}_{2} \} \nonumber \\
  & & \{ C_{2} \, (\Delta_{d} \, C_{3} \, \Delta^{\dagger}_{d}) \, C_{1u} \,
(\Delta \, C^{\dagger}_{3} \, \Delta^{\dagger}) \, C^{\dagger}_{2} \} \; .
\end{eqnarray} This matrix is not quite of the standard CKM form. However,
since the matrices $C_{i}$'s $\sim 1$, the matrix $U$ is not that different,
apart
from the placement of some phase factors.

\smallskip

Once one has explicit forms of the matrices $U$ and $D$, the ``$\lambda$''
expansion of these matrices and hence of $\tilde{M}_{u}$, $\tilde{M}_{d}$ is
accomplished by letting: \begin{equation}
\theta_{1d} \equiv \sum_{n=1}\alpha_{n}\lambda^{n} \;\; ; \;\; \theta_{2d}
\equiv \sum_{n=2}\beta_{n}\lambda^{n} \;\; ; \;\; \theta_{3d} \equiv
\sum_{n=4}\gamma_{n}\lambda^{n} \;\; . \label{theta_l}
\end{equation} The expansion of the $\theta_{iu}$ angles are fixed by
Eq.(\ref{theta_i}) in conjunction with the magnitudes of the $\theta_{i}$'s in
the CKM matrix, as specified in Eq.(\ref{CKM_No})\footnote{Notice, in
particular, that according to Eqs.(\ref{A}) and (\ref{sigma}), $s_{3} \equiv A
\sigma
\lambda^{3} \sim O(\lambda^{4})$, if one uses the central values for $A$ and
$\sigma$.} .
\medskip

Having written out the matrices $\tilde{M}_{u}$, $\tilde{M}_{d}$ according to
Eq.(\ref{MM}) (with $\tilde{M}_{u}^{diag}$, $\tilde{M}_{d}^{diag}$ given by
Eqs.(\ref{Mudiag}) and (\ref{Mddiag})) and expanded each matrix element in a
``$\lambda$'' expansion similar to Eq.(\ref{twogenexp}), one can follow a
procedure analogous to that in the two-generation case and infer which of the
mixing angle options give rise to natural mass matrices. The detailed
expressions for $\tilde{M}_{u}$ and $\tilde{M}_{d}$ are quite lengthy and not
tremendously illuminating, so we shall omit them and instead only present our
findings regarding the naturalness conditions. We find that for natural mass
patterns, we must require again that
\[ (1) \; \theta_{1d} \sim \lambda \; , \; \theta_{1u} \lsim \lambda^{2} \; ;
\] and, for the other angles, that one of the following options from each set
holds: \begin{eqnarray}
(2) & & \theta_{2u} \sim  \theta_{2d} \sim \lambda^{2} \; , \; \mbox{or} \; \;
\theta_{2u}  \sim  \lambda^{2} >> \theta_{2d}   \; , \; \mbox{or} \; \;
\theta_{2d}  \sim  \lambda^{2} >> \theta_{2u}  \; ; \; \nonumber \\
(3) & & \theta_{3u}  \sim  \theta_{3d} \sim \lambda^{4} \; , \; \mbox{or} \;
\;\theta_{3u}  \sim  \lambda^{4} >> \theta_{3d} \; , \; \mbox{or} \; \;
\theta_{3d}  \sim  \lambda^{4} >> \theta_{3u}   \; . \nonumber \\
& &  \label{option}
\end{eqnarray} The above conditions severely restricts the form of the mass
matrices to which they apply. As a result, the general expressions for these
matrices are readily obtained. (The detailed results and their discussion are
relegated to Appendix~A.) Here, as an example, we give a mass pattern with
$\theta_{1d} \sim \lambda \; , \; \theta_{1u} \sim \theta_{2d} \sim \lambda^{2}
\; , \; \theta_{2u} \sim \theta_{3u} \sim \lambda^{4} \; \mbox{and} \;
\theta_{3d} \sim \lambda^{5}$. This pattern corresponds to \begin{eqnarray}
\tilde{M}_{u} & \simeq & \left( \begin{array}{ccc}
 u_{11}\lambda^{7} &  u_{12}\lambda^{6}  &  e^{-i\delta_{u}}u_{13}\lambda^{4}
\\
 u_{12}\lambda^{6} &  u_{22}\lambda^{4}  &  u_{23}\lambda^{4} \\
e^{i\delta_{u}} u_{13}\lambda^{4} &  u_{23}\lambda^{4}  &  1
 \end{array} \right) \;\; ; \nonumber \\ \nonumber \\
\tilde{M}_{d} & \simeq &  \left( \begin{array}{ccc}
 d_{11}\lambda^{4} &  d_{12}\lambda^{3}  &  e^{-i\delta_{d}}d_{13} \lambda^{5}
\\
 d_{12}\lambda^{3} &  d_{22}\lambda^{2}  &  d_{23}\lambda^{2} \\
 e^{i\delta_{d}}d_{13} \lambda^{5} &  d_{23}\lambda^{2}  &  1
 \end{array} \right) \;\; , \label{matrix_pattern}
\end{eqnarray} where the real coefficients $u_{ij}$'s, $d_{ij}$'s are functions
of the following $O(1)$ parameters: the CKM parameters $A$, $\sigma$; the quark
mass ratios $\xi$'s; and the ``$\lambda$'' expansion coefficients $\{
\alpha_{1}, \beta_{2}, \gamma_{5} \}$. The coefficient $u_{13}$ and the phase
$\delta_{u}$ in addition also depend on the arbitrary phase parameter
$\delta_{d}$ from the matrix $\Delta_{d}$ (Eq.(\ref{D})) as well as on the CKM
phase $\delta$.

\smallskip

The principal goal of our construction is to allow us to extrapolate
LED-consistent natural mass patterns to some GUT scale where we can look for
 hints of ``new'' physics. Nevertheless, low energy (defined here as $\sim \,
m_{t}$) mass patterns such as the one discussed above are also interesting in
their own right. For instance, for the pattern given in
Eq.(\ref{matrix_pattern}), by appropriately choosing the signs of the quark
masses one can arrange to have
\begin{eqnarray}
\tilde{M}_{u} & \simeq & \left( \begin{array}{ccc}
 \xi_{ut}\lambda^{7} + (A\sigma)^{2}\lambda^{6} &  \alpha \xi_{ct}\lambda^{6}
&  - A\sigma e^{-i\delta}\lambda^{3}   \\
 \alpha \xi_{ct}\lambda^{6} &  \xi_{ct}\lambda^{4}  &  0 \\
 - A \sigma e^{i\delta}\lambda^{3}  & 0  &  1
 \end{array} \right) \; ; \nonumber \\ \nonumber \\
\tilde{M}_{d} & \simeq & \left( \begin{array}{ccc}
 0 &  \sqrt{\xi_{db} \xi_{sb}}\lambda^{3} &  0   \\
 \sqrt{\xi_{db} \xi_{sb}}\lambda^{3} &  \xi_{sb}\lambda^{2}  &  A\lambda^{2} \\
 0  &  A\lambda^{2}  &  1
 \end{array} \right)  \; ,
\end{eqnarray} with $\alpha =  \{\sqrt{\xi_{db} / \xi_{sb}} - 1 \} / \lambda
\simeq 0.12 $. This new pattern now has a large number of the much sought-after
``texture-zeros'': it has three exact ones to begin with, i.e.
$[\tilde{M}_{u}]_{23} \, , \, [\tilde{M}_{d}]_{11}$ and $[\tilde{M}_{d}]_{13}$;
and two more to the accuracy level $O(\lambda^{4})$ of the CKM matrix, i.e.
$[\tilde{M}_{u}]_{11}\, , \, [\tilde{M}_{u}]_{12} \sim O(\lambda^{7})$.
Moreover, this new pattern exhibits useful features commonly exploited in the
study of mass matrix patterns: sensible CKM and other ``predictions'' can come
about when one imposes equalities among matrix elements of approximately the
same order (e.g. demanding $|[\tilde{M}_{u}]_{13}| = [\tilde{M}_{u}]_{22}$
results in the prediction: $|V_{ub}| \simeq m_{c}/m_{t}$), or when one assigns
specific values (usually``0'') to certain (usually small) matrix elements (e.g.
 setting $[\tilde{M}_{u}]_{12} = 0$ results in the prediction: $\sin
\theta_{C}~= ~\sqrt{m_{d}/m_{s}}$).

\bigskip

\section{\normalsize \bf Potentially Successful GUT Scale Mass Patterns}
\setcounter{equation}{0}

\begin{flushleft}
\it{4.1.   A Pattern}
\end{flushleft}
\rm

Having constructed certain low energy natural mass patterns, one can then apply
RGE's to evolve these patterns to some high mass scales where global symmetries
originating from some GUT texture should become manifest, hopefully gaining
some useful insights. As a study case, in this paper we examine the evolution
of our
natural mass patterns in the MSSM theory. For simplicity, we consider only the
scenario where the VEV's of the Higgs coupled to u-quarks and d-quarks are
approximately equal, i.e.  $\tan \beta \simeq O(1)$. The relevant 1-loop
RGE's~\cite{DHR}\cite{OP} are: \begin{eqnarray}
\frac{d[h_{U}]_{ij}}{dt} & \simeq & \frac{1}{(4\pi)^{2}} \left\{
\rule{0in}{2ex} (3[h_{U}]^{2}_{33}-c_{k}g^{2}_{k})[h_{U}]_{ij} +
3[h_{U}]_{i3}[h_{U}]_{33}[h_{U}]_{3j} \right\} \; , \label{RGE_U} \\
\frac{d[h_{D}]_{ij}}{dt} & \simeq & \frac{1}{(4\pi)^{2}} \left\{
\rule{0in}{2ex} (3[h_{D}]^{2}_{33} + tr\{h^{2}_{E}\}
-c'_{k}g^{2}_{k})[h_{D}]_{ij} \right. \nonumber \\
& & \left. \hspace{1.3cm} + \, [h_{U}]_{i3}[h_{U}]_{33}[h_{D}]_{3j}
\rule{0in}{2ex} \right\}  \label{RGE_D}
\end{eqnarray} where the $g_{k}$'s are the three gauge couplings,
$c_{k}=(13/15,\,3,\,16/3)$, $c'_{k}=(7/15,\,3,\,16/3)$ and $h_{U} \, , \,
h_{D}$ and $h_{E}$ are the Yukawa coupling matrices for u-quarks, d-quarks and
leptons,\footnote{Although our analysis so far does not concern the lepton mass
matrix, one can still argue that it must be of the form which reflects the
lepton mass hierarchy, especially in light of the fact that one wishes to
implement the successsful Georgi-Jarlskog GUT mass relation~\cite{GJ}:
$m_{b}/m_{\tau}=3 m_{s}/m_{\mu}=m_{d}/3 m_{e} = 1$ at some point. It then
follows that $tr\{h^{2}_{E}\} \sim [h_{E}]^{2}_{33} \sim [h_{D}]^{2}_{33}$ , in
which case, the contribution of the term $tr\{h^{2}_{E}\}$ to the solution of
Eq.(\ref{RGE_D}) is very minimal.} respectively. To pursue our analysis
further, we need to solve the above equations to find the mass matrices at the
GUT scale
$m_{G} \simeq 10^{16}\,(GeV)$~\cite{DHR}. An input is then necessary at the
energy scale $m_{t}$. For definiteness, we assume as a boundary condition
$[h_{U}(m_{t})]_{33}=1$, although the general pattern of our result is largely
independent of this choice. For the concrete example of our mass pattern of
Eq.(\ref{matrix_pattern}), the solution of Eqs.(\ref{RGE_U}) and (\ref{RGE_D})
gives \small \begin{eqnarray}
\tilde{M}_{u}(m_{G}) & \simeq & \{0.82\} \left( \begin{array}{ccc}
 (\{0.61\}u_{11}+ \{0.08\}u^{2}_{13})\lambda^{7} & \{0.61\}u_{12}\lambda^{6}  &
 e^{-i\delta_{u}}u_{13}\lambda^{4}  \\
\{0.61\}u_{12}\lambda^{6} & \{0.61\}u_{22}\lambda^{4}  &  u_{23}\lambda^{4} \\
e^{i\delta_{u}} u_{13}\lambda^{4} & u_{23}\lambda^{4}  &  1
 \end{array} \right) ; \nonumber \\ \nonumber \\
\tilde{M}_{d}(m_{G}) & \simeq &  \{0.38\} \left( \begin{array}{ccc}
\{0.85\}d_{11}\lambda^{4} & \{0.85\}d_{12}\lambda^{3}  &
(\{0.85\}e^{-i\delta_{d}}d_{13} + \Delta_{13})\lambda^{5}\\
 \{0.85\}d_{12}\lambda^{3} &  \{0.85\}d_{22}\lambda^{2}  &
\{0.85\}d_{23}\lambda^{2} + \Delta_{23} \lambda^{4} \\
 e^{i\delta_{d}}d_{13} \lambda^{5} &  d_{23}\lambda^{2}  &  1
 \end{array} \right) \label{GUT_pattern} \, \nonumber \\
\end{eqnarray} \normalsize with $\Delta_{13}=\{0.68\}e^{-i\delta_{u}}u_{13}$
and $\Delta_{23}=\{0.15\}u_{23}$. The above result illustrates several general
features of the RG runnings of natural Hermitian mass patterns:

(1)As is obvious from the form of $\tilde{M}_{d}(m_{G})$, the Hermiticity of
the mass matrices is not strictly preserved by the RG evolution. However, the
extent to which Hermiticity is broken is relatively minor.

(2)Because of the hierarchy in the mass matrices at $m_{t}$, the RG runnings of
various mass matrix elements are quite different.

(3)This notwithstanding, the likely mass-matrix-element candidates for
``texture-zeros'' at $m_{G}$ are the same ones which are present at $m_{t}$.
These are the matrix elements of $O(\lambda^{4})$ or smaller.

These observations suggest a strategy on how to proceed in the search for GUT
patterns. First of all, since in practice it makes more sense to imagine the
mass matrices at the GUT scale to be Hermitian (or symmetric), one should
really reverse the procedure. Clearly, if one chooses the GUT pattern to be
Hermitian (e.g. by manipulating Eq.(\ref{GUT_pattern}) into its nearest
Hermitian form),
one should expect only modest deviation in the low energy mass matrices from
being perfectly LED-consistent. This is even less of a real problem since, as a
matter of fact, one can obtain LED-consistent mass matrices which are
non-Hermitian (see the more detailed discussion in Sec.~5). Secondly, one can
exploit the differences in the RG runnings of mass matrix elements to arrange
for possible equalities among them at the GUT scale. Finally, since
``texture-zeros'' track between high and low energy scales, one can look for
possible  ``texture-zeros'' of a GUT pattern in the ``texture-zeros'' or
``near-texture-zeros''of its corresponding low energy
pattern.\footnote{Equalities among matrix elements and ``texture-zeros' are
always desirable in GUT mass patterns in that they reduce the number of input
parameters and, as a result, enhance the predictive power of the patterns.}

\smallskip

For concreteness, it is useful to demonstrate our ideas with a specific
example. Let us consider again the mass pattern of Eq.(\ref{GUT_pattern}). By
choosing
the signs of the quark masses appropriately, one finds among other
possibilities, a potentially ``successful'' GUT mass pattern in which
\begin{eqnarray}
\tilde{M}_{u}(m_{G}) & = & \left( \begin{array}{ccc}
0 & C & B e^{-i\phi} \\
C & B & B \\
B e^{i\phi} & B & A
\end{array} \right) \;\; ; \label{pattern_u} \\ \nonumber \\
\tilde{M}_{d}(m_{G}) & = & \left( \begin{array}{ccc}
0 & F & 0 \\
F & E & E \\
0 & E & D
\end{array} \right) \;\; , \label{pattern_d}
\end{eqnarray} where the magnitudes of the parameters are : $A \sim O(1), \, B
\sim O(\lambda^{4}), \, C \sim O(\lambda^{6}) \; ; \; D \sim O(1), \, E \sim
O(\lambda^{2})$ and $F \, \sim O(\lambda^{3})$. Given this ansatz, one can then
try to find out whether it fits the LED. As we shall see below, some
straightforward computation shows that indeed it does.\footnote{Alternatively,
and in fact more efficiently, we could take as our starting point the Hermitian
mass pattern in Eq.(\ref{matrix_pattern}) to be our GUT pattern with all the
quark mass ratios and CKM parameters therein evaluated at the energy scale
$m_{G}$ (Appendix~B) and directly arrange for ``texture-zeros'' and equalities
among its matrix elements.}

\medskip

\begin{flushleft}
\it{4.2.   CKM Predictions}
\end{flushleft}
\rm

Running the RGE's backwards, at the energy scale $m_{t}$ the matrices of our
mass pattern become \begin{eqnarray}
\tilde{M}_{u} & \simeq & \left( \begin{array}{ccc}
-0.6{B'}^{2} & C' & B' e^{-i\phi} \\
C' & 1.6B' & B' \\
B' e^{i\phi} & B' & 1
\end{array} \right) \;\; ; \label{u_mt} \nonumber \\ \nonumber \\
\tilde{M}_{d} & \simeq & \left( \begin{array}{ccc}
0  & F'    & -0.18B'e^{-i\phi} \\
F' & 1.2E' & 1.2E' -0.18B' \\
0  & E'    & 1
\end{array} \right) \label{d_mt} \;\; .
\end{eqnarray} Notice that to our accuracy, only $\tilde{M}_{d}$ is slightly
non-Hermitian. For ease in the computations that follow, it is convenient to
specify  the approximate magnitudes of the various parameters above by defining
$B' \equiv b\lambda^{4}$, $C' \equiv c\lambda^{6}$; $E' \equiv e\lambda^{2}$,
and $F' \equiv f\lambda^{3}$. With these choices, one can relate the parameters
$b,c,e,f$ to the quark mass eigenvalues (with the signs of ``u'', ``c'' and
``d'' quark masses chosen to be negative) by solving the corresponding
eigenequations of the mass matrices, i.e. \begin{eqnarray*}
\det \left\{ \tilde{M}_{u} + \xi_{ct}\lambda^{4} \right\} & = & 0  \;\; , \\
\det \left\{ \tilde{M}_{d} \tilde{M}^{\dagger}_{d} - (\xi_{sb}\lambda^{2})^{2}
\right\} & = & 0  \;\; ...etc.
\end{eqnarray*} These computations give \begin{eqnarray*}
b & \simeq & - 0.6 \xi_{ct} \;\;\; , \\
c & \simeq & \pm \sqrt{\xi_{ct}(- 4.5 \xi_{ut} + 0.6 \xi^{2}_{ct}}) \;\;\; , \\
e & \simeq & 0.8 \xi_{sb} \;\;\; , \\
f & \simeq & \sqrt{\xi_{sb}\xi_{db}} \;\;\; .
\end{eqnarray*} With these values, one can further calculate the diagonalizing
unitary matrices  $U_{L}$, $D_{L}$ from the equations \begin{eqnarray*}
U^{\dagger}_{L} (\tilde{M}_{u}) U_{L} & = & \tilde{M}^{diag}_{u} \; , \\
D^{\dagger}_{L} (\tilde{M}_{d} \tilde{M}^{\dagger}_{d}) D_{L} & = & \{
\tilde{M}^{diag}_{d} \}^{2} \;\; .
\end{eqnarray*} The final result is \begin{eqnarray*}
[CKM] & = & U^{\dagger}_{L} D_{L} \\
& \simeq & \left( \begin{array}{ccc}
1 & \Delta_{12} & \Delta_{13} \\
-\Delta_{12} & 1 & \Delta_{23} \\
\Delta_{31}  & -\Delta_{23} & 1
\end{array} \right)
\end{eqnarray*} where \begin{eqnarray*}
\Delta_{12} & = & \sqrt{\xi_{db}/\xi_{sb}} \; \lambda \pm \sqrt{ -4.5
\xi_{ut}/\xi_{ct} + 0.6 \xi_{ct}} \; \lambda^{2} \;\; , \\
\Delta_{13} & = & 0.7 \xi_{ct} e^{-i\phi} \, \lambda^{4} \pm \xi_{sb}\sqrt{
-4.5 \xi_{ut}/\xi_{ct} + 0.6 \xi_{ct}} \; \lambda^{4} \;\; , \\
\Delta_{23} & = & \xi_{sb} \, \lambda^{2} + 0.7 \xi_{ct} \, \lambda^{4} \;\; ,
\\ \Delta_{31} & = & \sqrt{\xi_{db}\xi_{sb}} \; \lambda^{3} - 0.7 \xi_{ct}
e^{i\phi} \lambda^{4}  \;\; .
\end{eqnarray*} Comparing this matrix with the CKM matrix in the Wolfenstein
parametrization and denoting the absolute values of the quark masses as
$m_{q}$'s, one arrives at the following ``predictions'': \footnote{In terms of
the scaling parameter ``$r$'' defined in Appendix~B, the numerical factors in
these expressions correspond to $r \simeq 0.85$ (and hence $r^{2} \simeq 0.7,\,
r^{3} \simeq 0.6$).} \begin{eqnarray}
\sin\theta_{C} & \simeq &  \sqrt{\frac{m_{d}}{m_{s}}} \pm
\sqrt{-\frac{m_{u}}{m_{c}} + 0.6 \frac{m_{c}}{m_{t}}} \;\; , \nonumber \\
\nonumber \\
V_{cb} & \simeq &  \frac{m_{s}}{m_{b}} + 0.7 \frac{m_{c}}{m_{t}} \;\; ,
\nonumber \\ \nonumber \\
V_{ub} & \simeq & 0.7 \frac{m_{c}}{m_{t}} e^{-i\phi} \pm
\frac{m_{s}}{m_{b}}\sqrt{-\frac{m_{u}}{m_{c}} + 0.6 \frac{m_{c}}{m_{t}}} \;\; .
\label{CKM_pre}
\end{eqnarray} To check the soundness of these results, we choose the ``$+$''
sign in the above expressions and input various quark mass ratios. Although
these are not the only possible choices, we find that for  \[ \;\;\;\; \xi_{ct}
\simeq  1.55 \; ,\; \xi_{ut} \simeq 0.30 \; ; \; \xi_{sb} \simeq  0.73 \; , \;
\xi_{db} \simeq 0.66 \; , \]  we have a decent fit corresponding to the central
values of $\sin \theta_{C} \simeq 0.22$, $A \simeq 0.78$ and, $\sigma
e^{-i\delta} \simeq 0.31 e^{- i\phi} + 0.05$. Choosing $\phi = 90^{0}$ in the
last equation for example, gives the point $(0.05, \, 0.31)$ in the $\rho -
\eta$ plane, which is well within the known constraints (Eq.(\ref{rho_eta})).
We note also that the mass ratios above are in reasonable agreement with the
much
more restrictive light-quark-mass constraint relation of Eq.(\ref{light_q}).

\bigskip

\begin{flushleft}
\it{4.3.   Other GUT Patterns}
\end{flushleft}
\rm

By examining the general expressions for the GUT scale Hermitian matrices
$\tilde{M}_{u}$ and $\tilde{M}_{d}$ (Appendices~A and B), it is not difficult
to find other potentially interesting mass patterns. We list here four more
such
patterns which are slight variations of the one we discussed in detail above.
The first two have the same form for the $\tilde{M_{d}}$ matrices as the one in
Eq.(\ref{pattern_d}), i.e. \[ (1\, \& \,2) \hspace{1cm} \tilde{M}_{d}(m_{G}) =
\left( \begin{array}{ccc}
0 & F & 0 \\
F & E & E \\
0 & E & D
\end{array} \right) \] with $D \sim O(1), \, E \sim O(\lambda^{2})$ and $F
\sim O(\lambda^{3})$, but have somewhat different $\tilde{M_{u}}$'s:

\[ (1) \hspace{1cm} \tilde{M}_{u}(m_{G}) =  \left( \begin{array}{ccc}
C & 0 & B e^{-i\phi} \\
0 & B & B \\
B e^{i\phi} & B & A
\end{array} \right) \] with $A \sim O(1), \, B \sim O(\lambda^{4})$ and $C \sim
O(\lambda^{7})$;

\[ (2) \hspace{1cm} \tilde{M}_{u}(m_{G}) =  \left( \begin{array}{ccc}
C & C & B e^{-i\phi} \\
C & B & B \\
B e^{i\phi} & B & A
\end{array} \right) \] again with $A \sim O(1), \, B \sim O(\lambda^{4})$ and
$C \sim O(\lambda^{7})$. The remaining two patterns have a different
$\tilde{M_{d}}$ matrix which takes the form \[ (3\, \& \,4) \hspace{1cm}
\tilde{M}_{d}(m_{G}) = \left( \begin{array}{ccc}
0 & F & F e^{-i\phi}\\
F & E & E \\
F e^{i\phi} & E & D
\end{array} \right) \] with $D \sim O(1), \, E \sim O(\lambda^{2})$ and $F \sim
O(\lambda^{3})$; and the following $\tilde{M_{u}}$ matrices: \[ (3)
\hspace{1cm} \tilde{M}_{u}(m_{G}) =  \left( \begin{array}{ccc}
C & 0 & 0 \\
0 & B & B \\
0 & B & A
\end{array} \right) \] with $A \sim O(1), \, B \sim O(\lambda^{4})$ and $C \sim
O(\lambda^{7})$;

\[ (4) \hspace{1cm} \tilde{M}_{u}(m_{G}) =  \left( \begin{array}{ccc}
C & 0 & 0 \\
0 & 0 & B \\
0 & B & A
\end{array} \right) \] with $A \sim O(1), \, B \sim O(\lambda^{2})$ and $C \sim
O(\lambda^{7})$. The CKM ``predictions'' of the above GUT patterns are most
readily obtained in terms of various quark mass ratios and the parameter
``$r$'' defined in Appendix~B, by comparing the matrices of these patterns with
the
general results of Appendix~A applied at $m_{G}$ (see the example and comments
in Appendix~B for details). The relevant predictions for these GUT patterns are
tabulated below in Table~1.

\footnotesize
\begin{table}
\begin{tabular}{|c|c|c|c|c|}
\hline
 & (1) & (2) & (3) & (4) \\ \hline  & & & & \\
$\sin\theta_{C}$ &
$\sqrt{\dfrac{m_{d}}{m_{s}}}$ & $\sqrt{\dfrac{m_{d}}{m_{s}}} -
\dfrac{m_{u}}{m_{c}} + r^{3}\dfrac{m_{c}}{m_{t}}$ &
$\sqrt{\dfrac{m_{d}}{m_{s}}}$ & $\sqrt{\dfrac{m_{d}}{m_{s}}}$ \\ & & & & \\
\hline & & & &  \\
$V_{cb}$ & $\dfrac{m_{s}}{m_{b}} + r^{2} \dfrac{m_{c}}{m_{t}}$ &
$\dfrac{m_{s}}{m_{b}} + r^{2} \dfrac{m_{c}}{m_{t}}$ & $\dfrac{m_{s}}{m_{b}} +
r^{2} \dfrac{m_{c}}{m_{t}}$ &  $\sqrt{r \dfrac{m_{c}}{m_{t}}} -
\dfrac{m_{s}}{m_{b}}$ \\  & & & & \\ \hline & & & & \\
$V_{ub}$ & $r^{2} \dfrac{m_{c}}{m_{t}} e^{-i\phi}$ & $r^{2}
\dfrac{m_{c}}{m_{t}} e^{-i\phi} - \dfrac{m_{s}}{m_{b}} \left\{
\dfrac{m_{u}}{m_{c}} - r^{3} \dfrac{m_{c}}{m_{t}} \right\}$ &
$\sqrt{\dfrac{m_{d}m_{s}}{m^{2}_{b}}}  e^{-i\phi}$ & $ -
\sqrt{\dfrac{m_{d}m_{s}}{m^{2}_{b}}}  e^{-i\phi}$ \\ & & & & \\
\hline
\end{tabular}
\caption{GUT pattern predictions for CKM parameters}
\end{table}
\normalsize
\medskip

Several comments are in order at this point:

\smallskip

(1) Although the mass patterns listed here all contain ``texture-zeros'', we
have not tried to impose ``texture-zeros'' in all possible places. For example,
one could have in the matrix $\tilde{M}_{u}(m_{G})$ of pattern (1) an extra
``texture-zero'' by taking $C=0$. The resultant new pattern would, in addition
to its CKM ``predictions'', generate a GUT scale quark mass relation
corresponding to $m_{u}m_{t} \simeq r^{3} m^{2}_{c}$ which, according to
Eq.(\ref{Mudiag}), is actually allowed!  A systematic search for LED-consistent
patterns with the maximum number of ``texture-zeros'' has already been
thoroughly carried out in Ref.~\cite{RRR} where, specifically, a total of five
patterns with five ``texture-zeros'' were found and discussed in substantial
detail.

(2) Because of the specific mass matrix parametrization scheme we have chosen,
certain frequently-encountered  Hermitian mass patterns in the literature may
not seem transparent from the constructions of our natural mass matrices.
Still, in general these patterns can be related to our easily derivable
patterns
by some simple unitary transformations. For example,
 consider the following pattern
which can easily be arranged from our general results in
Appendix~A, \[
\tilde{M}_{u} \simeq \left( \begin{array}{ccc}
C & 0 & 0 \\
0 & B & 0 \\
0 & 0 & A
\end{array} \right) \;\; ; \;\; \tilde{M}_{d} \simeq \left( \begin{array}{ccc}
0 & F & G e^{-i\phi}\\
F & E & E' \\
G e^{i\phi} & E' & D
\end{array} \right) \; , \] where $A,\, D \sim O(1); \, E, \, E' \sim
O(\lambda^{2}); \, F  \sim O(\lambda^{3}); \, B,\, G \sim O(\lambda^{4})$ and
$C \sim O(\lambda^{7})$. With the various parameters carefully chosen, this
pattern can be transformed into a much more familiar-looking form~\cite{RRR} \[
\tilde{M'}_{u} \simeq \left( \begin{array}{ccc}
0 & C' & 0 \\
C' & B & 0 \\
0 & 0 & A
\end{array} \right) \;\; ; \;\; \tilde{M'}_{d} \simeq \left( \begin{array}{ccc}
0 & F' e^{-i\phi'} & 0 \\
F' e^{i\phi'} & E & E' \\
0 & E' & D
\end{array} \right) \] by a unitary matrix $T$ (i.e. $T \; \tilde{M}_{u,d} \;
T^{\dagger} \simeq \tilde{M'}_{u,d}$) with, \[ T \simeq \left(
\begin{array}{ccc}
e^{i\phi} & -\omega & 0 \\
\omega e^{i\phi} & 1 & 0 \\
0 & 0 & 1
\end{array} \right) \] where $\omega \equiv G/E' \sim O(\lambda^{2})$.

(3) If one wishes to incorporate the Georgi-Jarlskog mass relation\cite{GJ},
the corresponding lepton mass matrices of the GUT patterns listed in this
section
can be chosen in a straightforward manner. For instance, following
Ref.~\cite{DHR}, one can simply let, for patterns $1 \, \& \, 2$, \[ (1\, \&
\,2) \hspace{1cm} \tilde{M}_{l}(m_{G}) = \left( \begin{array}{ccc}
0 & F & 0 \\
F & -3E & E \\
0 & E & D
\end{array} \right) \;\; ; \] and similarly for patterns $3 \, \& \, 4$, one
can let \[ (3\, \& \,4) \hspace{1cm} \tilde{M}_{l}(m_{G}) = \left(
\begin{array}{ccc}
0 & F & F e^{-i\phi} \\
F & -3E & E \\
F e^{i\phi} & E & D
\end{array} \right) \;\; . \] With these matrices, it is easy to see that the
Georgi-Jarlskog relation results directly.

\bigskip

\section{\normalsize \bf Hermiticity Breakdown and Strong CP Complications}
\setcounter{equation}{0}

In our introductory discussion we described how to construct Hermitian mass
matrices from LED information. We must, however, face the fact that imagining
the mass matrices are Hermitian at the weak scale is not a very compelling
assumption. Indeed, as we have argued in the preceding sections it is much more
sensible to imagine that quark mass matrices are Hermitian (or symmetric) at
the GUT scale. When this is the case, the RG evolution definitely introduces
some
non-Hermitian (or non-symmetric) components at the weak scale. This was
illustrated in explicit detail in the example based on the mass matrices given
in Eqs.(\ref{pattern_u}) and (\ref{pattern_d}). Thus, to be realistic, we
should instead display a set of natural non-Hermitian weak scale mass matrices
constructed from the LED and then evolve these matrices to the GUT scale.

If one attempts this kind of a general construction from the LED without any
further constraints, one is immediately faced with considerable arbitraryness
and little progress appears possible. However, if one assumes that the
resulting weak scale matrices are only ``slightly'' non-Hermitian, because they
are
Hermitian at the GUT scale, then a general construction becomes feasible. In
fact, such a construction is really not necessary in the case of SUSY GUTs with
$\tan\beta \sim O(1)$. In this latter case one can simply detail how the CKM
parameters and the quark mass ratios evolve to the GUT scale. With these
parameters in hand one can directly construct natural Hermitian mass patterns
at the GUT scale. The resulting non-Hermitian mass matrices -- by construction
--
will be natural and reproduce the LED. This is basically the technique used to
deduce Table~1. The details of this procedure is further illustrated through an
example in Appendix~B.

The presence of non-Hermitian mass matrices at the weak scale, incidently,
raises the issue of strong CP violation. Because of the non-trivial
nature of the QCD vacuum~\cite{CDG}, the standard model is augmented by an
extra CP violating term involving the gluon field strength and its dual
 \[ \cal{L}_{\it Strong \, CP} = \it \frac{\alpha_{s}}{8\pi}
\,\overline{\theta} \, F_{a}^{\mu\nu} \tilde{F}_{a\mu\nu} \;\;. \] The
parameter $\overline{\theta}$ is a linear combination of a phase angle $\theta$
connected with the QCD vacuum
and another connected with the quark mass matrices~\cite{RDP_CP} \[
\overline{\theta} = \theta + \sum_{i=u,d} Arg\{\det M_{i}\} \;\; . \] One
knows, however, that this parameter must be extremely small ($\overline{\theta}
\leq
10^{-9}$)~\cite{edm}, so as to avoid being in conflict with the present bound
on the neutron electric dipole moment. Why should the QCD vacuum angle be so
precisely aligned as to cancel (or very nearly cancel) $Arg \{\det M_{u,d}\}$
is not known and constitutes the strong CP problem.

For the quark mass matrices we have been discussing, if we assume that at the
GUT scale these matrices are Hermitian then obviously \[ Arg \{ \det
M_{u,d}(m_{G}) \} = 0 \; . \]
However, as we have seen from our analysis, RG evolution induces
non-Hermiticity. Thus, starting with some Hermitian mass matrices $M_{u,d}$ at
the GUT scale, in general, these matrices become slightly non-Hermitian
at the scale of $m_{t}$. This is a direct consequence of the RGE's not being
Hermitian-conjugation invariant. In the 1-loop RGE (Eq.~\ref{RGE_D}), for
example, the term
$[h_{U}][h^{\dagger}_{U}][h_{D}]$ is responsible for this non-invariance.
Nevertheless, such a term is found to be insufficient to generate
\[ Arg \{ \det M_{u,d}(m_{t}) \} \neq 0
\; .  \] In general, however, one expects eventually that at sufficiently high
 order such a term will ensue from the RG evolution.

The actual order at which a non-zero value for $Arg \{ \det M_{u,d} \}$ at the
top scale appears depends on the underlying theory. For instance, in a globally
supersymmetric theory this mass matrix phase is never generated since it is not
renormalized~\cite{EFN}, while with the standard model it may first appear at
six loops in the Higgs sector, with an additional gauge boson loop~\cite{EG}.
In supersymmetric theories where SUSY is broken softly the actual contribution
depends on the breaking. In certain instances no mass matrix phase
appears~\cite{MS} but, in general, if there are non-vanishing elementary or
induced gluino masses one expects a phase to appear~\cite{DGH}. For instance,
with an explicit gluino mass, one induces a non-vanishing
$Arg \{ \det M_{u,d}(m_{t}) \}$ at two-loops~\cite{DGH}.

It is quite possible that with the right underlying theory, imposing \linebreak
$Arg \{ \det M_{u,d}(m_{G}) \}=0$ at the GUT scale suffices to guarantee that
 \linebreak $Arg \{ \det M_{u,d}(m_{t}) \}$ is much below $10^{-9}$. However,
 this still does not solve the strong CP problem unless, somehow,
$\overline{\theta}$ vanishes at $m_{G}$ (which certainly is not sufficiently
guaranteed by just having $Arg \{ \det M_{u,d}(m_{G}) \} = 0$). These
additional observations indicate perhaps compellingly the necessity of having
some
dynamical strong-CP-removal mechanism, conceivably by imposing a $U(1)_{PQ}$
symmetry~\cite{PQ}.

\bigskip

\section{\normalsize \bf Conclusions}
\setcounter{equation}{0}

Interesting patterns of quark masses are surely signals of ``new'' physics. The
task of searching for them therefore can be very rewarding. In order to conduct
these searches more effectively, we have suggested in this paper the idea of
natural mass matrices as an organizing principle. This idea, along with the
efficient mass-matrix-parametrization scheme we have described, allows a
procedure whereby one can systematically input low energy data to construct
viable GUT patterns.  Encouragingly, this procedure has produced a rather small
set of ``working'' mass patterns and our preliminary work in extrapolating
these patterns to GUT scales has generated some interesting possiblities. We
have
discussed, specifically, one such application in the context of SUSY GUTs and
some potentially successful GUT mass patterns were readily found. Although we
do not particularly wish to assign too much significance to these mass patterns
and their predictions, such examples do indicate the usefulness of our
approach. An
important future task is to perform a more systematic and complete
investigation, with different RGE boundary conditions and perhaps different
matter contents.

\bigskip

\begin{flushleft}
\bf{Acknowledgements}
\end{flushleft}
\rm

One of us (RDP) would like to thank S. Raby for a useful discussion. We are
also grateful to X. G. He for a helpful comment on the strong CP problem,
which corrected an earlier misconception. This work is supported in part by
the Department of Energy under Grant No. FG03-91ER40662.

\bigskip

\appendix
\section{\normalsize \bf General Expressions for Natural Mass Matrices}

Adopting the parametrization scheme we have developed in Sec.~3.2, we arrive at
the following general expressions for natural mass matrices, incorporating all
naturalness requirements (Eq.(\ref{option})): \begin{eqnarray}
\tilde{M}_{u} & \simeq & \left( \begin{array}{ccc}
 u_{11}\lambda^{7} &  u_{12}\lambda^{6}  &  u_{13}\lambda^{4}  \\
 u^{*}_{12}\lambda^{6} &  u_{22}\lambda^{4}  &  u_{23}\lambda^{2} \\
 u^{*}_{13}\lambda^{4} &  u_{23}\lambda^{2}  &  1
 \end{array} \right) \nonumber \\ \nonumber \\
& + &  \left( \begin{array}{ccc}
 O(\lambda^{9}) &  O(\lambda^{8})  &   O(\lambda^{6})  \\
 O(\lambda^{8}) &  O(\lambda^{6})  &   O(\lambda^{6}) \\
 O(\lambda^{6}) &  O(\lambda^{6})  &   O(\lambda^{4})
 \end{array} \right) \ \;\; ;  \label{matrix_GPu} \\ \nonumber \\
\tilde{M}_{d} & \simeq &  \left( \begin{array}{ccc}
 d_{11}\lambda^{4} &  d_{12}\lambda^{3}  &  d_{13} \lambda^{4}  \\
 d_{12}\lambda^{3} &  d_{22}\lambda^{2}  &  d_{23}\lambda^{2} \\
 d^{*}_{13} \lambda^{4} &  d_{23}\lambda^{2}  &  1
 \end{array} \right) \nonumber \\ \nonumber \\
& + &  \left( \begin{array}{ccc}
 O(\lambda^{6}) &  O(\lambda^{5})  &   O(\lambda^{6})  \\
 O(\lambda^{5}) &  O(\lambda^{4})  &   O(\lambda^{4}) \\
 O(\lambda^{6}) &  O(\lambda^{4})  &   O(\lambda^{4})  \end{array} \right) \;\;
. \label{matrix_GPd}
\end{eqnarray} Here, \begin{eqnarray}
u_{11} & = & \xi_{ut} + \{ \alpha^{2}_{u}\xi_{ct} + | \gamma_{u}
e^{-i\delta_{d}} - \alpha_{u}A  + \Lambda (e^{-i\delta_{d}} - e^{-i\delta})
|^{2} \} \lambda \;\; , \nonumber \\
u_{12} & = & \alpha_{u}\xi_{ct} + \beta_{u}[\gamma_{u}e^{-i\delta_{d}}-
\alpha_{u} A + \Lambda (e^{-i\delta_{d}} - e^{-i\delta}) ] \;\; , \nonumber \\
u_{13} & = & \gamma_{u}e^{-i\delta_{d}} -\alpha_{u}A  + \Lambda
(e^{-i\delta_{d}} - e^{-i\delta}) \;\; , \nonumber \\
u_{22} & = & \xi_{ct} + \beta^{2}_{u} \;\; , \nonumber \\
u_{23} & = & \beta_{u} \;\; ; \nonumber \\
d_{11} & = & \xi_{db} + \alpha^{2}_{d} \xi_{sb} \;\; , \nonumber \\
d_{12} & = & \alpha_{d}\xi_{sb} \;\; , \nonumber \\
d_{13} & = & \gamma_{d} e^{-i\delta_{d}} - \alpha_{d}\beta_{d}\xi_{sb}\lambda
\;\; , \nonumber \\
d_{22} & = & \xi_{sb} \;\; , \nonumber \\
d_{23} & = & \beta_{d} \;\; . \label{ud}
\end{eqnarray} The various parameters in the above equations are as follows:
$\lambda, A, \Lambda (\equiv \sigma A \lambda^{-1})$ and $\delta$ are the CKM
parameters (Eq.(\ref{CKM_W})); $\xi$'s are the quark mass ratios
(Eqs.(\ref{Mudiag}) and ~(\ref{Mddiag})); $\delta_{d}$ is a free input phase
parameter (Eq.(\ref{D})); and finally $\alpha$'s, $\beta$'s and $\gamma$'s are
input parameters defined from the ``$\lambda$'' expansion coefficients of the
$\theta_{u,d}$'s (Sec.~3.2): \[ \begin{array}{ll}
\sin \theta_{1d} \equiv \alpha_{d}\lambda \;\; , & \sin \theta_{1u} \equiv
\alpha_{u}\lambda^{2} \;\; ; \\
\sin \theta_{2d} \equiv \beta_{d}\lambda^{2} \;\; , & \sin \theta_{2u} \equiv
\beta_{u}\lambda^{2} \;\; ; \\
\sin \theta_{3d} \equiv \gamma_{d}\lambda^{4} \;\; , & \sin \theta_{3u} \equiv
\gamma_{u}\lambda^{4} \;\; .
\end{array} \] The magnitudes of these last parameters are specified below in
accordance with our naturalness condition (Eq.(\ref{option})): \begin{eqnarray}
\alpha_{d} \sim O(1) \;,\; \alpha_{u} \lsim O(1) \;, && \mbox{with} \;\;\;
\alpha_{d} -\alpha_{u}\lambda = 1 + O(\lambda^{3}) \;\; ; \nonumber \\
\beta_{d} \lsim O(1) \;,\; \beta_{u} \lsim O(1) \;, && \mbox{with} \;\;\;
\beta_{d} -\beta_{u} = A + O(\lambda^{4}) \;\; ;\nonumber \\
\gamma_{d} \lsim O(1) \;,\; \gamma_{u} \lsim O(1) \;, && \mbox{with} \;\;\;
\gamma_{d} -\gamma_{u} = \Lambda + O(\lambda^{8}) \;\; . \label{constraints}
\end{eqnarray}

\medskip

The general expressions summarized here are particularly useful for the purpose
of arranging interesting mass patterns. Specifically, ``texture-zeros'' and
equalities among mass matrix elements can, whenever possible, be rather
conveniently imposed by adjusting the parameters $\alpha$'s, $\beta$'s and
$\gamma$'s, subject to the constraints in Eq.(\ref{constraints}). Furthermore,
CKM and other ``predictions'' then ensue when the aforementioned constraints
overspecify these parameters.

\bigskip

\section{\normalsize \bf Mass Ratios, CKM Parameters and Constructions of
Hermitian Mass Patterns at the GUT Scale}
\setcounter{equation}{0}

The relevant formulas for calculating the RG scaling of mass ratios and CKM
parameters are derived in Ref.~\cite{OP}\cite{Babu}. Here, we give only a brief
summary of the results. For the SUSY GUT case we are considering (Sec.~4.1)
where $\tan \beta \simeq O(1)$, one finds the following simple RG scaling
relations:

\medskip

-- Mass ratios
\begin{eqnarray}
\xi_{ct}(m_{G}) \simeq r^{3} \, \xi_{ct} &,& \xi_{ut}(m_{G}) \simeq r^{3} \,
\xi_{ut} \; ; \nonumber \\
\xi_{sb}(m_{G}) \simeq r \, \xi_{sb}  &,& \xi_{db}(m_{G}) \simeq r \, \xi_{db}
\; . \label{GUT_mass}
\end{eqnarray}

\medskip

-- CKM parameters
\begin{eqnarray}
\lambda(m_{G}) & \simeq & \lambda \; ,  \nonumber \\
A(m_{G}) & \simeq & r \, A \; ,  \nonumber \\
\sigma(m_{G}) & \simeq & \sigma \; . \label{GUT_CKM}
\end{eqnarray} The scaling parameter $r$ in these relations is defined by
\begin{equation}
r = e^{-\frac{1}{(4\pi)^{2}} \int_{0}^{\ln(m_{G}/m_{t})} {[h_{U}(\mu)]^{2}_{33}
dt}} \;\;\;\; (t \equiv \ln \{\mu/m_{t}\})
\end{equation} which, based on Eq.(\ref{RGE_U}) and the 1-loop RGE's for the
gauge couplings\footnote{These are, $dg_{i}/dt=b_{i}g^{3}_{i}/16\pi^{2} \;\;
(i=1,2,3)$ with $b_{i}=(33/5,\,1,\,-3)$.}, can also be expressed as
\begin{equation}
r = \left\{ \frac{[h_{U}(m_{G})]_{33}}{[h_{U}(m_{t})]_{33}} \right\}^{-1/6}
\{\eta(m_{G})\}^{1/12}  \label{r2}
\end{equation} with \begin{equation}
\left\{\frac{[h_{U}(m_{G})]_{33}}{[h_{U}(m_{t})]_{33}}\right\} \simeq
\{\eta(m_{G})\}^{1/2} \, \left\{
1-\frac{3}{4\pi^{2}}\,[h_{U}(m_{t})]^{2}_{33}\,I(m_{G}) \right\}^{-1/2}  \;,
\label{h33}
\end{equation} \[ \eta(\mu) \equiv
\prod^{i}\{g_{i}(m_{t})/g_{i}(\mu)\}^{2c_{i}/b_{i}} \;\; \] and \[ I(\mu)
\equiv \int_{0}^{\ln(\mu/m_{t})} \eta(\mu) \, dt \;\;.\]
To obtain a numerical value for $r$, we input $g^{2}_{i}(m_{t})/4\pi \simeq
(0.017,\,0.033,\,0.100)$ as values for the gauge couplings (at
$m_{t}$)~\footnote{These numbers were also used to produce
Eq.(\ref{GUT_pattern}). They correspond to a set of values for the gauge
couplings (at $m_{Z}$) used as inputs in Ref.~\cite{DHR} where, solving the
1-loop RGE's with these inputs, the three gauge couplings were found to merge
at $m_{G} \simeq 10^{16} \, (GeV)$.} along with the boundary condition
$[h_{U}(m_{t})]_{33}=1$  into the above results, and we find $r \simeq
0.85$.\footnote{Notice from Eq.(\ref{h33}) that the solution for
$[h_{U}(m_{G})]_{33}$ depends rather sensitively on the choice of the boundary
condition. However, while still important, this dependence is comparatively
speaking much milder for $r$.}
\smallskip

The general expressions for natural Hermitian mass matrices at the GUT scale
can be gotten by substituting the mass ratios and CKM parameters  evaluated at
$m_{G}$ in Eqs.(\ref{GUT_mass}) and (\ref{GUT_CKM}) into the expressions given
in Appendix~A. For illustrative purposes, we ``derive'' a somewhat generic
Hermitian GUT pattern and its CKM ``predictions'' below.

\smallskip

Choosing in Eqs.(\ref{matrix_GPu}) - (\ref{ud}) the parameters
$\delta_{d}=\delta, \, \beta_{u} \sim O(\lambda^{2})$ and demanding $u_{11}, \,
d_{11} \lsim O(\lambda^{2})$, one arrives at a mass pattern which can be
written as \begin{eqnarray}
\tilde{M}_{u} & \simeq & \left( \begin{array}{ccc}
0 & C & y_{u}B e^{-i\phi_{u}} \\
C & B & x_{u}B \\
y_{u}B e^{i\phi_{u}} & x_{u}B & A
\end{array} \right) \;\; , \nonumber \\
\tilde{M}_{d} & \simeq & \left( \begin{array}{ccc}
0 & F & y_{d}F e^{-i\phi_{d}}\\
F & E & x_{d}E \\
y_{d}F e^{i\phi_{d}} & x_{d}E & D
\end{array} \right) \label{adj_pattern}
\end{eqnarray} where $A,\, D \sim O(1); \, E \sim O(\lambda^{2}); \, F  \sim
O(\lambda^{3}); \, B \sim O(\lambda^{4}); \, C \sim O(\lambda^{6})$; and $x$'s,
$y$'s are adjustable parameters which are constrained only by the naturalness
requirement.

\smallskip

Mapping the matrix elements in Eq.(\ref{adj_pattern}) onto those in
Eq.(\ref{ud}), one can immediately establish the following: \begin{eqnarray*}
\beta_{u} \simeq x_{u}\xi_{ct}\lambda^{2}  & , & \beta_{d} \simeq x_{d}\xi_{sb}
\;\; ; \\
-\alpha_{u}A + \gamma_{u} e^{-i\delta} & \simeq & y_{u}\xi_{ct} e^{-i\phi_{u}}
\;\;, \\
-\alpha_{d}\beta_{d}\xi_{sb}\lambda + \gamma_{d} e^{-i\delta} & \simeq &
y_{d}\alpha_{d}\xi_{sb}\lambda^{-1} e^{-i\phi_{d}} \;\; ; \\
\xi_{ut} + \alpha^{2}_{u}\xi_{ct}\lambda + (y_{u}\xi_{ct})^{2}\lambda & \simeq
& 0 \;\; , \\
\xi_{db} + \alpha^{2}_{d}\xi_{sb} & \simeq & 0 \;\; .
\end{eqnarray*} Next, solving for the parameters $\alpha$'s, $\beta$'s and
$\gamma$'s from the above equations and subsequently applying the CKM
constraint relations given in Eq.(\ref{constraints}), one has,
\begin{eqnarray*}
1 & \simeq & \sqrt{-\xi_{db}/\xi_{sb}} \pm \sqrt{-\xi_{ut}/ (\xi_{ct}\lambda) -
y^{2}_{u}\xi_{ct}} \; \lambda  \;\; , \\
A & \simeq & x_{d}\xi_{sb} -  x_{u}\xi_{ct} \, \lambda^{2} \;\; , \\
\Lambda e^{-i\delta} & \simeq & y_{d} \xi_{sb} \sqrt{-\xi_{db}/\xi_{sb}}
e^{-i\phi_{d}} \, \lambda^{-1} - y_{u}\xi_{ct} e^{-i\phi_{u}}  \\
& & \pm A \sqrt{-\xi_{ut}/ (\xi_{ct}\lambda) - y^{2}_{u}\xi_{ct}} + x_{d}
\sqrt{-\xi_{db}\xi^{3}_{sb}} \; \lambda  \;\; .
\end{eqnarray*} Finally, in the above equations, if one keeps only the
significant terms and takes into account the RG scaling of mass ratios and CKM
parameters (i.e.  Eqs.(\ref{GUT_mass}) and (\ref{GUT_CKM})), one sees that the
GUT pattern of Eq.(\ref{adj_pattern}) has the following CKM ``predictions'':
\begin{eqnarray*}
\sin\theta_{C} & \simeq &
\sqrt{-\frac{m_{d}}{m_{s}}} \pm \sqrt{-\frac{m_{u}}{m_{c}} -
y^{2}_{u}r^{3}\frac{m_{c}}{m_{t}}} \;\; , \\
V_{cb} & \simeq & x_{d}\frac{m_{s}}{m_{b}} - x_{u}r^{2} \dfrac{m_{c}}{m_{t}}
\;\; , \\
V_{ub} & \simeq & y_{d}\frac{m_{s}}{m_{b}}\sqrt{-\frac{m_{d}}{m_{s}}}
e^{-i\phi_{d}} - y^{2}_{u}r^{2} \frac{m_{c}}{m_{t}} e^{-i\phi_{u}} \pm x_{d}
\frac{m_{s}}{m_{b}} \sqrt{-\frac{m_{u}}{m_{c}} -
y^{2}_{u}r^{3}\frac{m_{c}}{m_{t}}} \;\; .
\end{eqnarray*} The signs of the quark masses above have yet to be chosen as
either ``$+$'' or ``$-$'', depending on which choice is more sensible and
gives better agreement with experimental measurements of the CKM parameters.

\newpage

\pagebreak


\begin{thebibliography}{AAAAAA}
\bibitem{FRITZ} H. Fritzsch, Phys. Lett. 70B (1977) 436; 73B (1978) 317
\bibitem{DHR} S. Dimopoulos, L. Hall and S. Raby, Phys. Rev. D45 (1992) 4192;
G. Anderson, S. Raby, S. Dimopoulos, and L. J. Hall, Phys. Rev. D47 (1993) 3702
\bibitem{ARDHS} G. Anderson, S. Raby, S. Dimopoulos, L. J. Hall and G. D.
Starkman, Phys. Rev. D49 (1994) 3660; K. S. Babu, R. N. Mohapatra, Phys. Rev.
Lett. 74 (1995) 2418
\bibitem{PECCEI_talk} R. D. Peccei, talk presented at the International
Symposium on Particle Theory and Phenomenology, Iowa State University, Ames,
Iowa, May 1995, UCLA/95/TEP/30
\bibitem{AN} C. H. Albright, S. Nandi, Phys. Rev. Lett. 73 (1994) 930
\bibitem{RRR} P. Ramond, R.G. Roberts and G.G. Ross, Nucl. Phys. B406 (1993) 19
\bibitem{PDG} Particle Data Group: L. Montanet {\it et al.}, Phys. Rev. D50
(1994) 1173
\bibitem{WOLF} L. Wolfenstein, Phys. Rev. Lett. 51 (1983) 1945
\bibitem{Stone} S. Stone, in the Proceedings of the 1994 DPF Conference,
Albuquerque, N. Mexico, ed. S. Seidel (World Scientific, Singapore, 1995)
\bibitem{PW} R. D. Peccei and K. Wang, Phys. Lett. B349 (1995) 220
\bibitem{CDF} S. Abachi {\it et al.} (D0 Collaboration), Phys. Rev. Lett. 74
(1995) 2632; F. Abe {\it et al.} (CDF Collaboration), Phys. Rev. Lett. 74
(1995) 2626
\bibitem{GL} J. Gasser and H. Leutwyler, Phys. Rep. 87 (1982) 77
\bibitem{KM} D. Kaplan and A. Manohar, Phys. Rev. Lett. 56 (1986) 2004
\bibitem{Leut} H. Leutwyler, Nucl. Phys. B337 (1990) 108
\bibitem{OP} M. Olechowski and S. Pokorski, Phys. Lett. B257 (1991) 388
\bibitem{GJ} H. Georgi and C. Jarlskog, Phys. Lett. B86 (1979) 297
\bibitem{CDG} C. G. Callan, R. F. Dashen and D. J. Gross, Phys. Lett. 63B
(1976) 334
\bibitem{RDP_CP} R. D. Peccei, The Strong CP Problem, in {\bf CP Violation},
ed. C.~Jarlskog (World Scientific, Singapore, 1989)
\bibitem{edm} Particle Data Group: J. J. Hernandez {\it et al.}, Phys. Lett.
B239, (1990) 1
\bibitem{EFN} J. Ellis, S. Ferrara and D. V. Nanopoulos, Phys. Lett. B114
(1982) 231
\bibitem{EG} J. Ellis and M. K. Gaillard, Nucl. Phys. B150 (1979) 141
\bibitem{MS} R. N. Mohapatra, S. Ouvry and G. Senjanovi\'c, Phys. Lett. B126
(1983) 329; D. Chang and R. N. Mohapatra, Phys. Rev. D32 (1985) 293
\bibitem{DGH} M. Dugan, B. Grinstein and L. Hall, Nucl. Phys. B255 (1985) 413
\bibitem{PQ} R. D. Peccei and H. R. Quinn, Phys. Rev. D16 (1977) 1791
\bibitem{Babu} K. S. Babu, Z. Phys. C35 (1987) 69
\end{thebibliography}
\end{document}